\documentclass[
    ,final            % use final for the camera ready runs
  ]
  {aipproc}

\usepackage{graphicx}
\usepackage{amssymb}
\usepackage{amsmath}
\usepackage{amsfonts}
\usepackage{array,hhline,dcolumn} % Better table handling
\usepackage{epstopdf}

\usepackage{algorithm,algorithmic}
\usepackage{url}
\usepackage{color}
\usepackage{tikz}
\usepackage{multirow}

\newcommand{\opal}{\textsc{OPAL}}
\newcommand {\htp}{$\text{H}_2^+$}

\layoutstyle{6x9}

\begin{document}

\title{Modelling challenges of the high power cyclotrons for the DAE$\delta$ALUS project}


\author{A. Adelmann for the DAE$\delta$ALUS collaboration}{
address={Paul Scherrer Institut, CH-5232 Villigen, Switzerland}
}

%\author{J.R. Alonso}{
%address={Department of Physics, Massachusetts Institute of Technology}
%}
%
%\author{W.A. Barletta}{
%address={Department of Physics, Massachusetts Institute of Technology}
%}
%
%
%\author{L. Calabretta}{
%address={Istituto Nazionale di Fisica Nucleare - LNS}
%}
%
%\author{A. Calanna}{
%address={Department of Physics, Massachusetts Institute of Technology}
%}
%
%\author{D. Campo}{
%address={Department of Physics, Massachusetts Institute of Technology}
%}
%
%
%\author{J. J. Yang}{
%address={Department of Physics, Massachusetts Institute of Technology \& China Institute of Atomic Energy, Beijing, 102413, China}
%}

\begin{abstract}
 Design studies, for accelerator modules based on an injector cyclotron and a superconducting ring cyclotron able to accelerate H$_2^+$ molecules, are presented. H$_2^+$ molecules are stripped by a foil creating a proton beam, with a maximum energy of 800 MeV and a beam power of 8 MW (CW). This beam would be sent to a beam dump where neutrinos would be produced from pion and muon decays at rest for the Decay At rest Experiment for $\delta_{CP}$ At the Laboratory for Underground Science - DAE$\delta$ALUS. We are discussing the advantage of H$_2^+$ molecules for acceleration and present precise beam dynamics simulations w.r.t. extraction and beam losses. In general, beam losses are one of the most challenging parts in such a high power cyclotron design and must be addressed very early on in the design. We are also addressing H$_2^+$ dissociation and the stripping process, two other characteristic challenges in the DAE$\delta$ALUS design.
\end{abstract}

\maketitle

\section{Introduction}
\begin{figure}
  \includegraphics[width=0.58\linewidth]{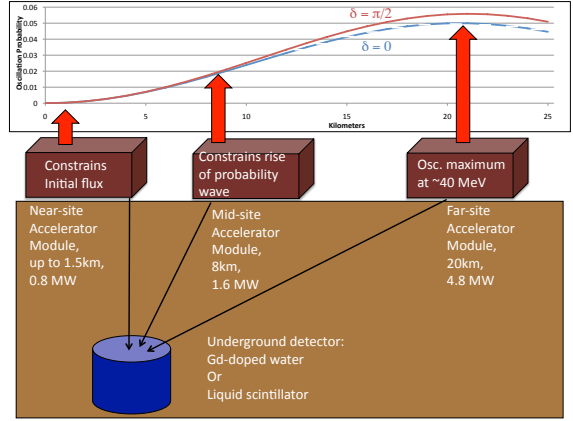}
  \caption{Schematic of the layout of DAE$\delta$ALUS accelerator modules. The powers at the respective modules, are average values based on a 20\% duty cycle.} \label{fig:1}
\end{figure}

In Fig. \ref{fig:1} schematically the layout for the DAE$\delta$ALUS experiment is shown.\ Three sites, at 1.5 km, 8 km and 20 km establish the necessary conditions for $\nu$-oscillations with expected high sensitivity to the CP-violating term $\delta$ \cite{schol}.  The power levels of 0.8 MW at the near site, 1.6 MW at intermediate and 4.8 MW at the far site, are calculated to yield data rates commensurate with a 10-year experiment, and were designed to be complementary with the planned LBNE experiment that proposed a new beamline from Fermilab to a 200 kTon water-Cherenkov counter situated at the 4850 level of the Sanford Underground laboratory in Lead, South Dakota (a 1000 km baseline). The neutrino sources are each isotropic, so net flux at the detector varies as $1/r^2$; however the signal grows approximately with distance squared, moderating the flux loss. 
Though the large water-Cherenkov counter at Homestake is no longer on the near-term planning boards, the possibility exists of siting at other suitable large detectors hosting long-baseline experiments (LENA, MEMPHYS, or HyperK, as examples), and the benefits remain for conducting both the short and long-baseline programs at the same time. 
\subsection{REQUIREMENTS FOR DAE$\delta$ALUS NEUTRINO SOURCES}
Optimum proton energy for DAE$\delta$ALUS pion production is around 800 MeV.  This energy is comfortably within the delta resonance range; is low enough to minimize decay-in-flight contamination and hence to minimize background of electron antineutrinos in the primary neutrino flux from unabsorbed $\pi^{-}$. Approximately 40\% of the time all sources must be off to obtain accurate background measurements.  If each site runs for 20\% of the time then the instantaneous beam current (and beam power) must be a factor of 5 higher than the average.  For the near site to achieve 0.8 MW, average current must be 1 mA, peak current 5 mA.  Peak current at the 8 km site must be 10 mA and 30 mA at the far (20 km) site. 
Beam quality and time structure of the beam on target are immaterial, however beam losses must be kept exceedingly low, of the order of parts in $10^{-4}$, to not preclude hands-on maintenance of the accelerators and their components. In this paper we are addressing the beam dynamic study for the mid-site cyclotrons (ref. to Fig.\ \ref{fig:1}).
The injector cyclotron (DIC) is a four-sector compact machine, which accelerates a beam of \htp up to 60 MeV/amu. 
The beam is then extracted by an electrostatic deflector and is transported and injected into an eight-sector superconducting ring cyclotron (DSRC), 
in which  the beam is accelerated to 800 MeV/amu by four single-gap cavities.
Two stripper foils are used to extract two proton beams at the same time from the ring cyclotron.  Main parameters are summarized in Tab.\ \ref{tab:cycs}.
\begin{table}[htb] 
\caption{\label{tab:cycs} Key parameters of the mid-site cyclotrons, both with harmonic 6 and 4 cavities (DIC, double-gap \& DSRC single-gap }
\begin{tabular}{lcccccccccc}
\hline \hline
& type & kin. energy &  avg. power & avg. field &sector&  turn\\
&  & (MeV/amu) & (MW) & (T) & no.  & no.\\
\hline
DIC & normal & 0.035\dots60 & 0.12 & 0.95\dots1.17 &  4 &107 \\
DSRC & superconducting&  60\dots800&1.6 & 1.06\dots1.88 & 8 & 401 \\
\hline \hline
\end{tabular}
\end{table}

\section{ADDRESSING SOME OF THE  BEAM DYNAMICS CHALLENGES}
The space-charge effects are studied quantitatively by self-consistent 3D models implemented in the code Object Oriented Parallel Accelerator Library (\opal)\ \cite{opal}. The beam dynamics model is described in detail in \cite{PhysRevSTAB.13.064201}. In the case of the DIC we conducted a modeling campaign similar to \cite{PhysRevSTAB.14.054402} for the PSI facility, currently marking the intensity frontier of CW proton drivers. For the DSRC, we implemented a simple stripper model into \opal\ in order to study the complex extraction trajectories of the stripped protons. 
\begin{figure}
  \includegraphics[height=.25\textheight]{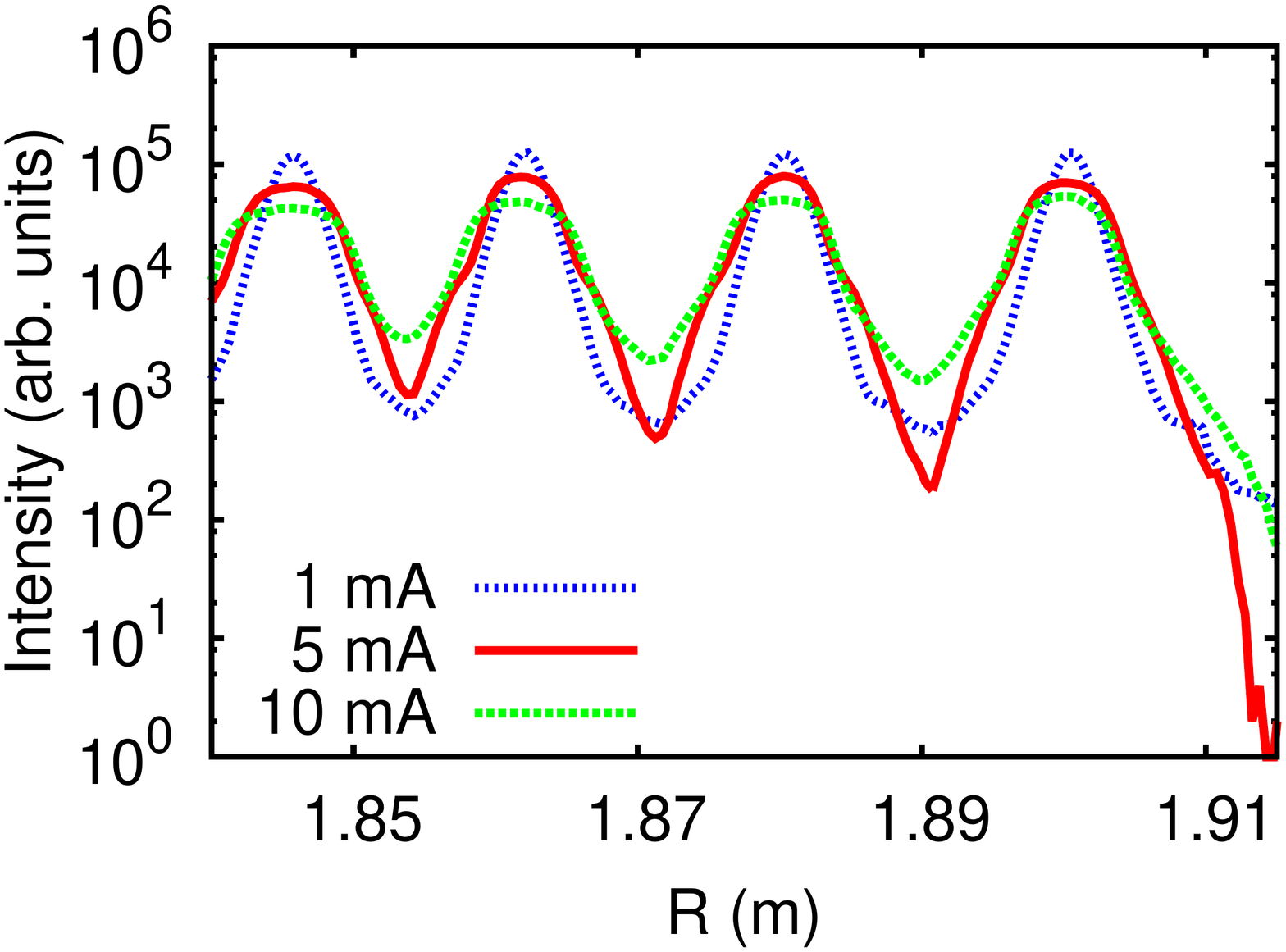}
  \includegraphics[height=.25\textheight]{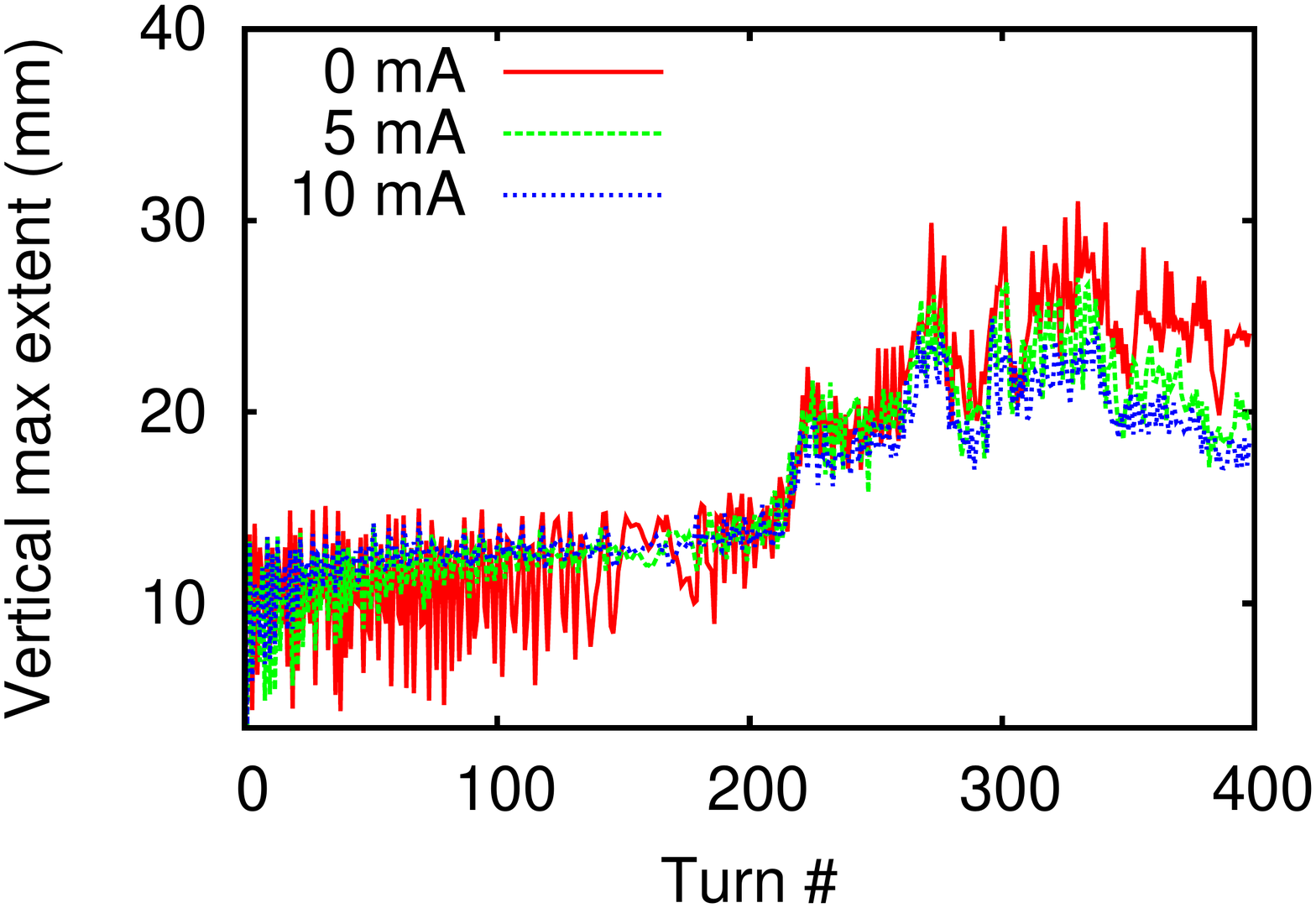}
  \caption{Left: the radial profile of the last 4 turns at the center on the valley for the different beam current with the initial phase width of $20^\circ$; right: the maximum beam extent in the vertical direction over the full acceleration cycle.} \label{fig:2}
\end{figure}
Important results concerning the DIC, namely the losses at the extraction septum, is shown in the left part of Fig.\ \ref{fig:2}. From this results and with the experience of the PSI facility, we can conclude that the single-turn extraction scheme for the DIC is feasible. In the case of the DSRC, a stripper extraction scheme does not require separated turns, hence we are more concerned with energy spread and vertical stability of the beam. On the right part of Fig.\ \ref{fig:2}, we see the maximum beam extend in vertical direction, which is well in the 80 (mm) aperture of the DSRC and hence does
not pose problems for our design. For more details we refer to \cite{EricePaper}.



\bibliographystyle{aipproc}   % if natbib is available

\bibliography{51-adelmann}

\IfFileExists{\jobname.bbl}{}
 {\typeout{}
  \typeout{******************************************}
  \typeout{** Please run "bibtex \jobname" to optain}
  \typeout{** the bibliography and then re-run LaTeX}
  \typeout{** twice to fix the references!}
  \typeout{******************************************}
  \typeout{}
 }

\end{document}